\documentclass[graybox]{svmult}

\usepackage{type1cm}        
\usepackage{caption}
\usepackage{makeidx}         
\usepackage{graphicx}        
\usepackage{subcaption}                             
\usepackage{multicol}  
\usepackage{wrapfig,lipsum}
\usepackage[bottom]{footmisc}

\usepackage{float}

\usepackage{newtxtext}       %
\usepackage{newtxmath}       

\usepackage{nccmath}

\makeindex             

\begin{document}

\title*{A 3D Kinetic Distribution that Yields Observed Plasma Density in Inner Van Allen  Belt }
\author{Snehanshu Maiti and Harishankar Ramachandran}
\institute{Snehanshu Maiti \at Indian Institute of Technology Madras, Chennai, \email{snehanshu.maiti@gmail.com}
}
%
%
\maketitle

\abstract*{A steady-state distribution is obtained that approximately yields the observed plasma density profile of the inner Van Allen radiation belt. The model assumes a collision-less, magnetized plasma with zero electric field present. The inner Van Allen belt consists a plasma comprising of high energy protons and relativistic electrons. The particle trajectories are obtained from the collision-less Lorentz Force equation for different initial distributions. The resulting steady state distributions obtained after  particles lost to the loss cone are eliminated and is used to generate the density profile. The distribution's dependence on  energy E and magnetic moment $\mu$ are adjusted to make the density profile agree with observations. For a distribution that is a function of energy times a function of magnetic moment the calculation leads to the desired type of density profile. The kinetic distribution and the type of density profile obtained are presented.}

\abstract{A steady-state distribution is obtained that approximately yields the observed plasma density profile of the inner Van Allen radiation belt. The model assumes a collision-less, magnetized plasma with zero electric field present. The inner Van Allen belt consists a plasma comprising of high energy protons and relativistic electrons. The particle trajectories are obtained from the collision-less Lorentz Force equation for different initial distributions. The resulting steady-state distributions obtained after  particles lost to the loss cone are eliminated and are used to generate the density profile. The distribution's dependence on  energy E and magnetic moment $\mu$ is adjusted to make the density profile agree with observations. For a distribution that is a function of energy times a function of magnetic moment, the calculation leads to the desired type of density profile. The kinetic distribution and the type of density profile obtained are presented.}

\section{Introduction}
\label{sec:1}
The inner Van Allen radiation belt exists approximately from an altitude of 1000-6000 km, (0.2-2) $R_E$ above the Earth's surface and contains electrons in the range of hundreds of KeV and energetic protons exceeding 100 MeV, trapped by the strong  geomagnetic field (relative to the outer belts) in the region [1]. The plasma is collision-less in nature and experiences Lorentz force within the magnetosphere. The particles are confined in a magnetic mirror and undergo gyro-motion, bounce motion and drift motion around the earth.

The plasma density observed is given by a commonly used form of an exponential (oxygen)  plus a power law (hydrogen) [2].
\begin{ceqn}
\begin{align}
   n(r)= {n_O} {e^{-(r-R_I)/h}} +{n_H}r^{-1}
\end{align}
\end{ceqn}  			   
where r is the geocentric radius, $n_O = 10^5 cm^{-3}$ is the density of oxygen,
$n_H = 10^3 cm^{-3}$  is the density of hydrogen, h = 400 km is the scale height and $R_I =1.0314 R_E$, where $R_E$ is radius of earth. 
\begin{figure}[H]
\centering
  \includegraphics[width=0.5\linewidth]{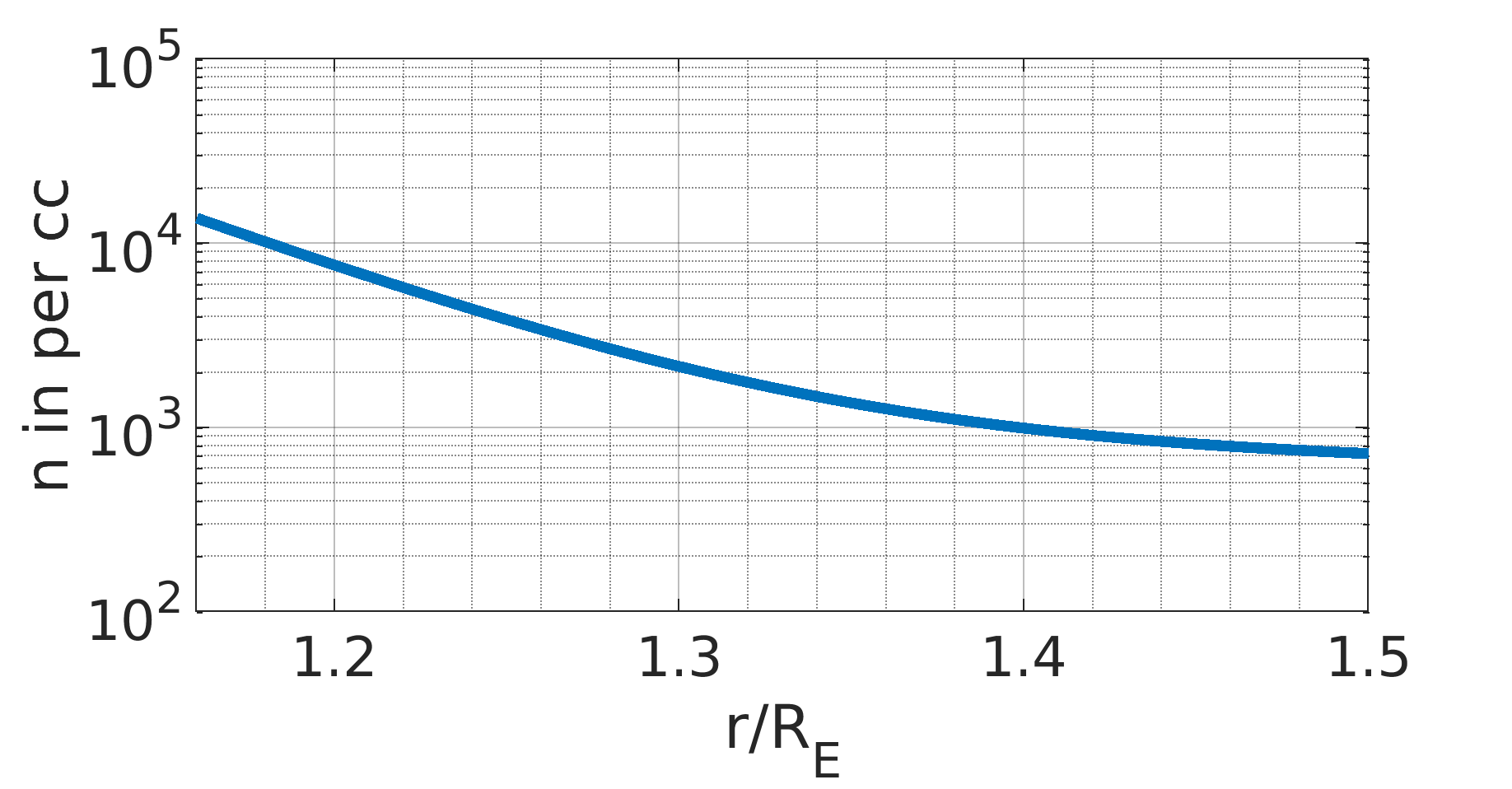}
  \vspace{-1em}
 \caption{Observed density profile of the radiation belt with altitude at the equator.}
  \label{fig:boat1}
\end{figure}
The current aim is to find a steady-state kinetic distribution of the particles which yields a density distribution closely resembling the observed density function at the equator as in Figure 1. We predict an initial f(E,$\mu$) distribution analytically and use the same to do numerical test particle simulations by considering different combinations of perpendicular and parallel energy to arrive at and closely mimic the density profile of the radiation belt and compare the two methods.
 
A numerical model of the radiation belt established can help study the effect of disturbances in the belt due to whislers, ULF waves, solar activities, seismo-electric activities, etc. and resulting particle precipitations from the belt to make predictions for IITM nano-satellite mission [3]. 
\section{ Analytical Model}
\label{sec:2}

The density distribution can be found out by integrating the phase space distribution function of the particles over the velocity space. This is written in  polar coordinate as follows and integrated over the entire $\phi$ direction from 0 to $2\pi$.

\begin{ceqn}
\begin{align}
 n=\int \int \int f(v)v^2 \sin\theta dv d\theta d\phi ,\qquad n=2\pi \int \int f(v)v^2 \sin\theta dv d\theta 
 \end{align}
\end{ceqn}		 	 				      
The above equation can be next converted to a function of  E and  $\mu$  by replacing Eq.(2) with following Eqs.(3) to obtain Eqs.(4-6).
\begin{ceqn}
\begin{align}
E = \frac{1}{2}mv^2 ,  dE=mvdv , \quad
v_\perp = v \sin\theta,  dv_\perp=v \cos\theta d\theta , \quad
  \mu=\frac {mv_\perp ^ 2}{2B} , d\mu=\frac {mv_\perp dv_\perp}{B}
\end{align}
\end{ceqn} 
\begin{ceqn}
\begin{align}
n=\frac{2\pi}{m}\int& \int f(E,\mu) dEv\sin\theta d\theta \\
n=\frac{2\pi}{m}\int \int f(E,\mu) dE \frac{v_\perp}{v\cos\theta}& dv_\perp  ,\quad
n=\frac{2\pi}{m}\int \int f(E,\mu) dE \frac{v_\perp}{v_\parallel} dv_\perp 
\end{align}
\end{ceqn} 				         
\begin{ceqn}
\begin{align}
  n(s)=\frac{2\pi B(s)}{m^2}\int^\infty_0 \int^\frac{E}{B}_0 \frac{f(E,\mu) dE d\mu}{\sqrt{E-\mu B(s)}}
\end{align}
\end{ceqn}        			   
The plasma in the Van Allen belts is collisionless and hence chosen to have a near Maxwellian distribution of thermal energy as follows.
A simple distribution function f($E,\mu$ ) is chosen such that,
\begin{ceqn}
\begin{align}
  f(E)=E e ^  \frac{-E}{KT} ,\qquad f(\mu) = \mu e ^{-\mu}
\end{align}
\end{ceqn} 	          			    
The density profile from the above predicted distribution is presented in Fig. 6(b) as a comparison with observed density and numerical simulations.

\section{ Numerical Model}
\label{sec:3}


The inner radiation belt being located between L shell 1.5-2.5, the geomagnetic field line at L shell 1.5 is considered for this simulation.  A dipole model of the Earth's  magnetic field is considered here which is a first order approximation of the rather complex true Earth's magnetic field and holds good for lower L shells [1].
In a dipole model, the geocentric radi $r$, the geomagnetic latitude $\theta$ considered northwards from the equator and the arc length s along L shell are related as:    
\begin{ceqn}
\begin{align}
  r = Lcos^2\theta , \quad ds^2=dr^2 +(rd\theta)^2 , \quad d\theta=\frac{ds}{L\sqrt{sin^22\theta+cos^4\theta}}
\end{align}
\end{ceqn}								   
A polar plot of L shell 1.5 is presented in Fig.2(a) using Eqns.8. The 's' coordinate system is  considered along the L shell with origin at the equator. In the s
coordinates, $s_{max}$ (towards the poles) represents the value corresponding to a radial
distance r = h +$R_E$ or an altitude of h = 1000 kms above the surface of the earth
(radius = $R_E$) and where the magnetosphere ends. The value of smax is 5044 kms in
the s coordinates.
The dipole model magnetic field strength in  polar coordinates is given below and used in Lorentz force equation to simulate particle trajectories.
\begin{ceqn}
\begin{align}
  B(r)=B_0(\frac{R_E}{r})^3 \sqrt{(1+3\sin^2\theta)}
\end{align}
\end{ceqn}

Particles are distributed uniformly in position along the 1D s coordinate initially as obtained in Fig.2(b) using random number generator algorithm. 
\begin{figure}[H]
\centering
\includegraphics[width=0.7\linewidth]{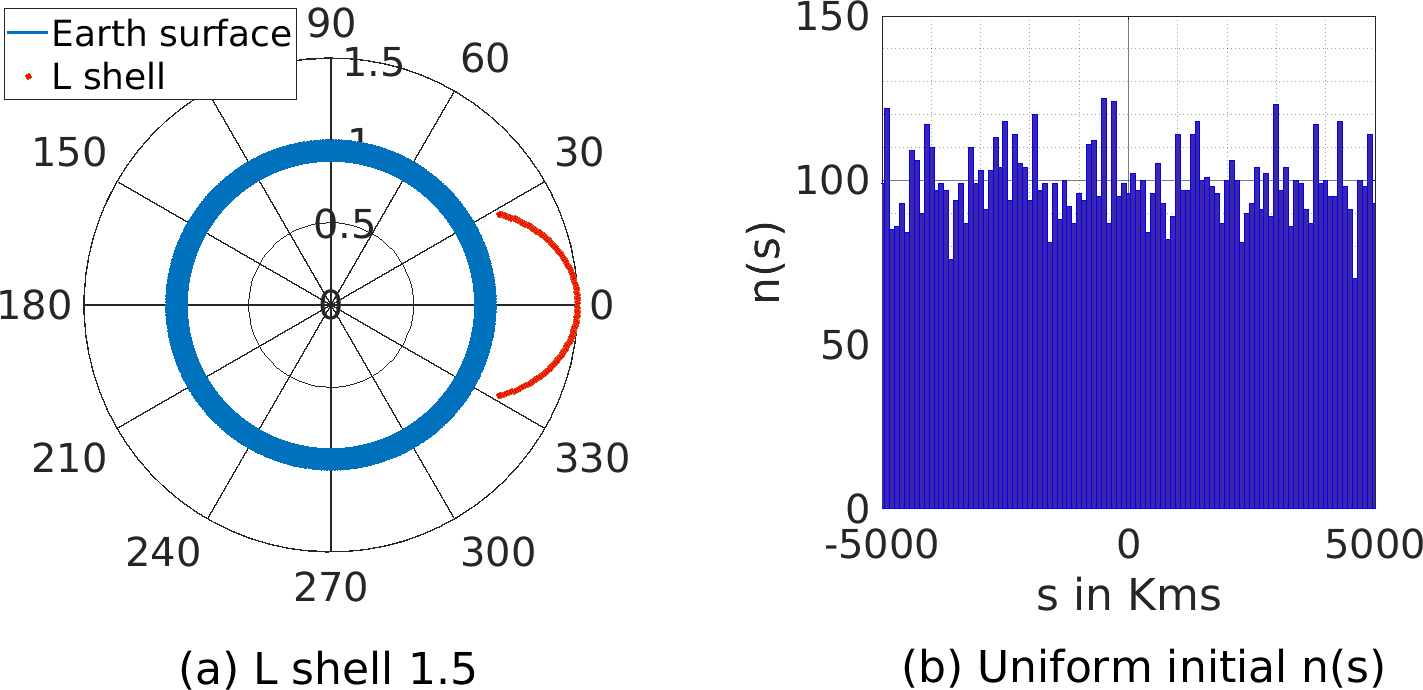}

\caption{(a) shows a polar plot of the lower magnetosphere Lshell 1.5 and (b) shows particles distributed uniformly along the L shell.}
\end{figure}
This 1D position distribution of the particles is transformed into 2D polar (r,$\theta$) coordinates using Eqns.(8) and  is further converted into 3D Cartesian coordinates x, y, z as : $x=r\cos\theta$, $y=0$ and $z=r \sin \theta$.\\

Next an initial f(E,$\mu$) is chosen. $f(E)$ is chosen as the analytical model. f($\mu$) is $f(E)/ (\alpha B_{max})$ where $E_{\perp}=\mu B$ is varied as $\alpha= \frac{E}{\mu B}$ for different cases to obtain different density functions. This initial f(E,$\mu $) is presented in Fig 3 for $\alpha=20$ and is converted into 3D velocity space below. \\

\begin{figure}[H]
\centering
  \includegraphics[width=1\textwidth]{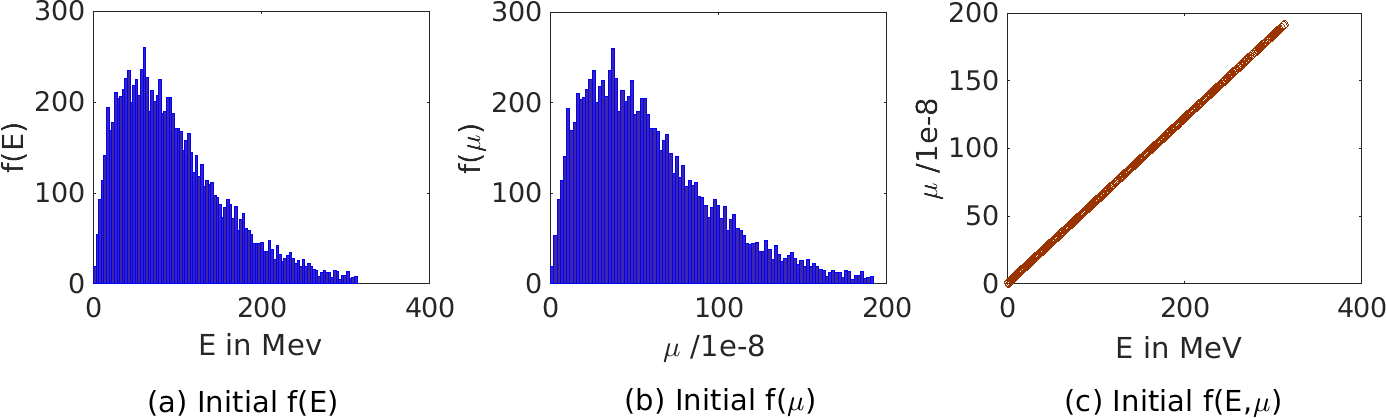}
  \caption{Initial energy and magnetic moment distribution of 10000 protons.}
  \label{fig:boat1}
\end{figure}

The initial parallel and perpendicular velocity distribution can be resolved into $v_x$,$v_y$ and $v_z$ as, 
\begin{ceqn}
\begin{align}
  v_\perp=\frac{2 \mu B}{m}  , \quad
   v_x=v_\perp \cos \phi , v_y=v_\perp \sin \phi , v_z= \sqrt{v^2-{v_x}^2-{v_y}^2}
\end{align}
\end{ceqn} 

Test particle simulations are run with 10,000 protons for 2 secs with the above initial distributions  and the trajectory of the particles obey Lorentz force. The Runge-Kutta 4 method is used to solve the ODE. $T_{gyro}=\frac{2\pi m}{qB}=2000\mu s$. So, the timestep of $10\mu s$ is chosen which satisfies Nyquist criteria and captures the particle trajectory accurately. The particles have an average bounce period of 0.2 secs [1] and hence attain the steady state in Fig. 6(a) in very short time as seen in n(t). Many  particles are lost due to smaller initial pitch angle and rest of the particles attains steady state. \\


The density profile, n(s) is presented in Fig.4 for different values of $\alpha$=1 and 20 and also if density is obtained only from f(E) without varying f(E,$\mu$). This shows a best fit to observed density for $\alpha=20$.
\begin{figure}[H]
\centering
\includegraphics[width=1\linewidth]{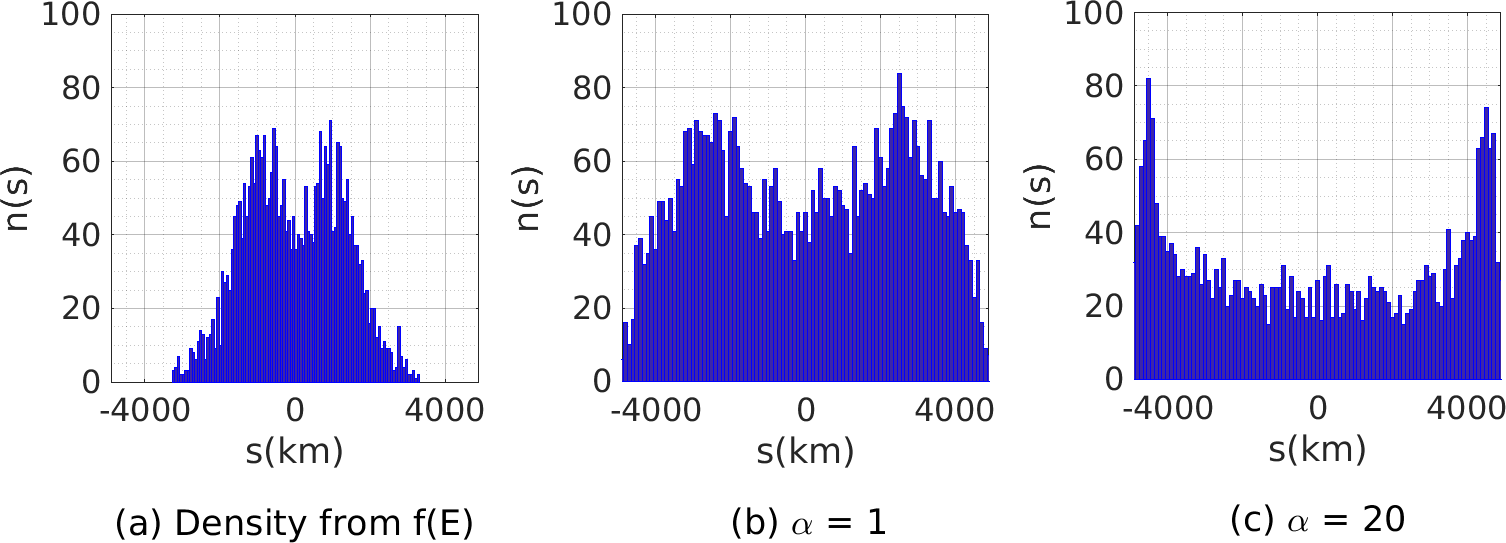}
\caption{Density distribution for different values of $\alpha$ shows best fit for $\alpha$=20.}
\end{figure}
The E-$\mu$ distributions at steady state is presented in Fig. 5 for $\alpha$=20. The lost particles (in loss cone) clearly separates out from the trapped particles in orbit in Fig.5(c) when $\mu$ is observed at their bounce points. 

\begin{figure}[H]
\centering
  \includegraphics[width=1\textwidth]{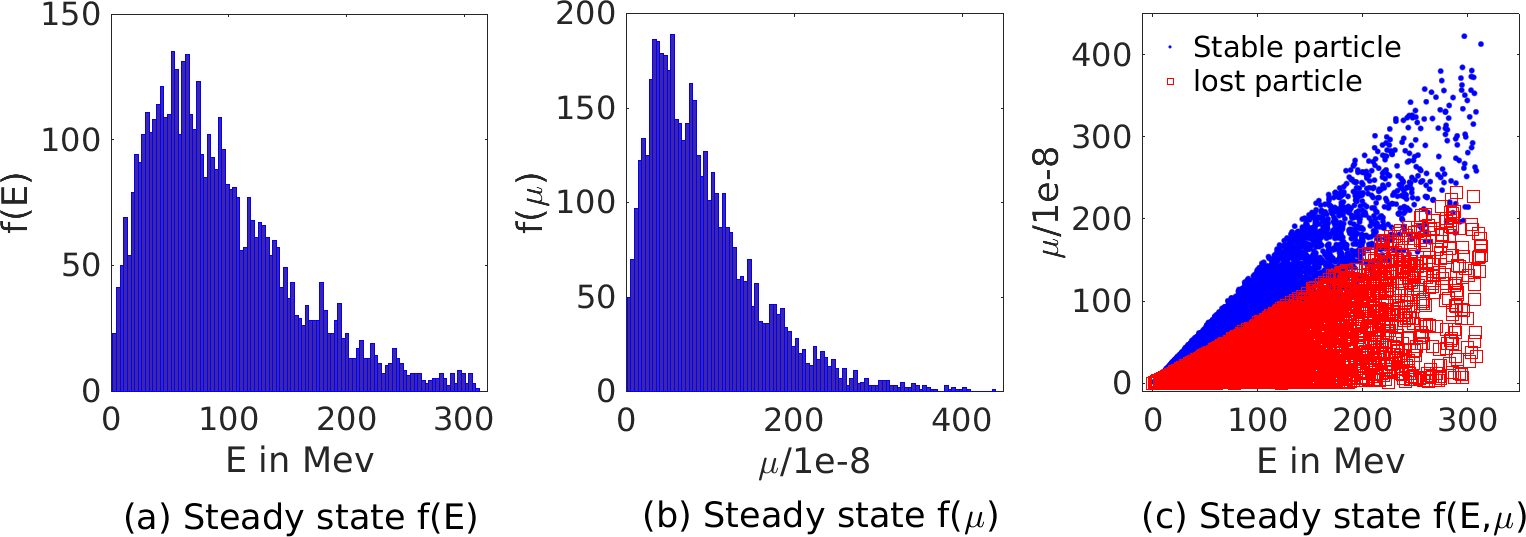}
  \caption{Energy and magnetic moment distribution of 10000 protons at steadystate.}
  \label{fig:boat1}
\end{figure}
\section{Conclusion}
\label{sec:3}

\begin{figure}[H]
\centering
\includegraphics[width=1\linewidth]{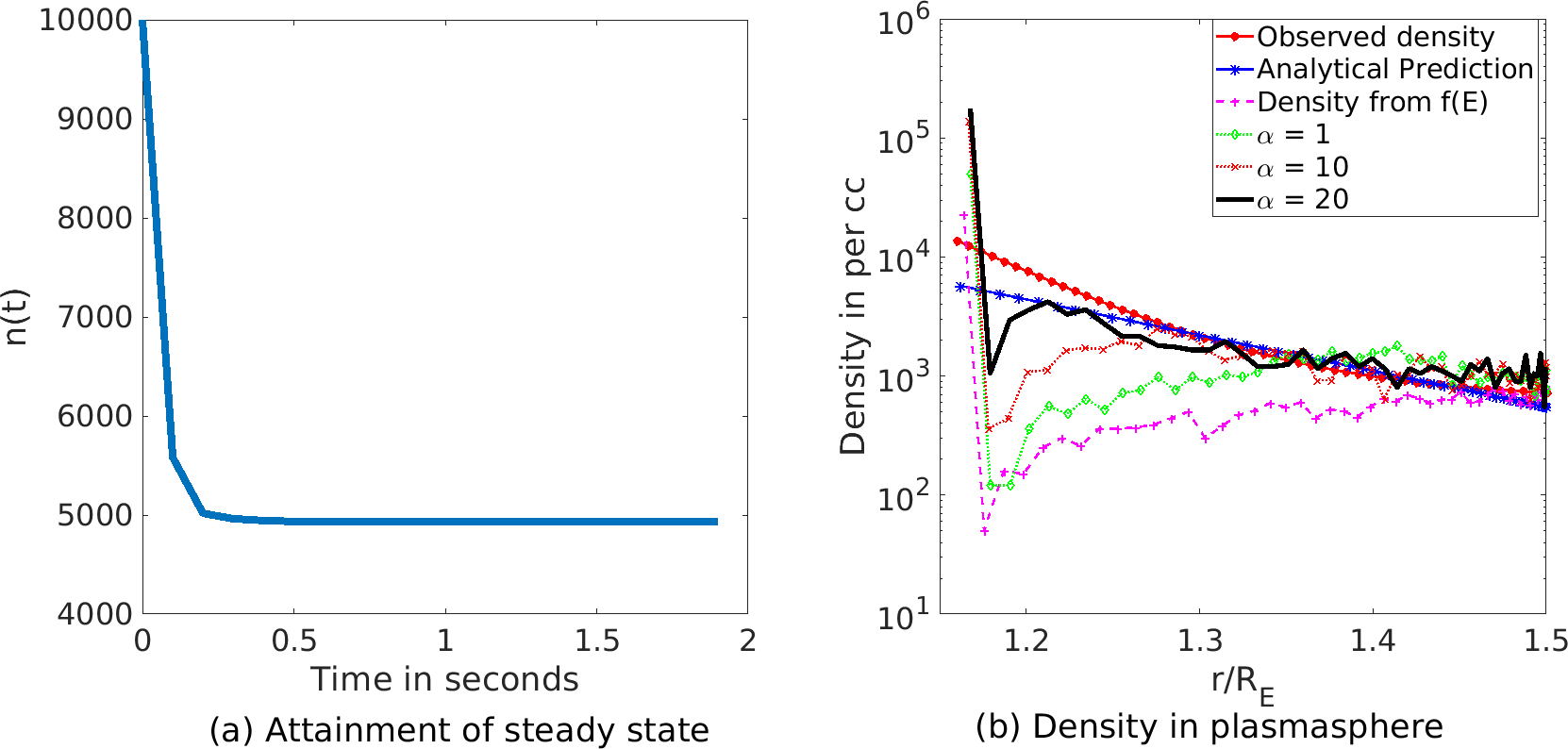}
\caption{(a) shows the attainment of steadystate of particles through n(t) and (b) shows the density profiles obtained for different values of $\alpha$ }\label{dc}
\end{figure}

Fig.~\ref{dc}(b) compares the observed, analytically predicted and numerically obtained densities.

An $\alpha=20$ represents best the observed density. The analytical method gives approximate particle guiding centre trajectory whereas the numerical simulation takes care of true gyromotion. Hence the loss cone could be properly studied using the current numerical simulations.

\end{document}